\documentclass[aps,twocolumn,superscriptaddress,amssymb]{revtex4}

\usepackage{amsfonts,amssymb,amsmath}
\usepackage{graphicx}
\usepackage{dsfont}
\usepackage{soul}
\usepackage{mathtools}
\usepackage{placeins}
\RequirePackage[
  colorlinks,
  linkcolor=blue,
  urlcolor=blue,
  citecolor=blue,
  plainpages=false,
  pdfpagelabels,
]{hyperref}

\graphicspath{{./figs/}}

\newcommand{\ket}[1]{\ensuremath{\left|#1\right\rangle}}

\newcommand{\mean}[1]{\ensuremath{\left\langle #1 \right\rangle}}

\newcommand{\tcr}[1]{#1}

\renewcommand{\st}[1]{}

\begin{document}
    \title{Variational quantum algorithm with information sharing}

\author{Chris N. Self}
\email{cns08@ic.ac.uk}
\affiliation{Blackett Laboratory, Imperial College London, London SW7 2AZ, United Kingdom}

\author{Kiran E. Khosla}
\affiliation{Blackett Laboratory, Imperial College London, London SW7 2AZ, United Kingdom}

\author{Alistair W. R. Smith}
\affiliation{Blackett Laboratory, Imperial College London, London SW7 2AZ, United Kingdom}

\author{Fr\'{e}d\'{e}ric Sauvage}
\affiliation{Blackett Laboratory, Imperial College London, London SW7 2AZ, United Kingdom}

\author{Peter D. Haynes}
\affiliation{Blackett Laboratory, Imperial College London, London SW7 2AZ, United Kingdom}
\affiliation{Department of Materials, Imperial College London, London SW7 2AZ, United Kingdom}

\author{Johannes Knolle}
\affiliation{Blackett Laboratory, Imperial College London, London SW7 2AZ, United Kingdom}
\affiliation{Department of Physics TQM, Technische Universit\"{a}t M\"{u}nchen, James-Franck-Stra{\ss}e 1, D-85748 Garching, Germany}
\affiliation{Munich Center for Quantum Science and Technology (MCQST), 80799 Munich, Germany}

\author{Florian Mintert}
\affiliation{Blackett Laboratory, Imperial College London, London SW7 2AZ, United Kingdom}


\author{M. S. Kim}
\affiliation{Blackett Laboratory, Imperial College London, London SW7 2AZ, United Kingdom}

\date{\today}

\begin{abstract}
We introduce \tcr{\st{a new }an} optimisation method for variational quantum algorithms and experimentally demonstrate a 100-fold improvement in efficiency compared to naive implementations. The effectiveness of our approach is shown by obtaining multi-dimensional energy surfaces for small molecules and a spin model. Our method solves related variational problems in parallel by exploiting the global nature of Bayesian optimisation and sharing information between different optimisers. Parallelisation makes our method ideally suited to next generation of variational problems with many physical degrees of freedom. This addresses a key challenge in scaling-up quantum algorithms towards demonstrating quantum advantage for problems of real-world interest. 
\end{abstract}

\maketitle

\section{Introduction}

Rapid developments in quantum computing hardware~\cite{jurcevic2020demonstration,arute2019quantum,pino2020demonstration} have led to an explosion of interest in near-term applications~\cite{preskill_quantum_2018,benedetti2019parameterized,smith_simulating_2019,mcardle2020quantum,bauer2020quantum}. Though current devices are remarkable feats of engineering, their current coherence times and gate fidelities exclude running general quantum algorithms such as Shor's factorisation, Grover search or quantum phase estimation. Nevertheless, it is hoped that Variational Quantum Algorithms (VQA's) will be able to demonstrate a quantum advantage on Noisy Intermediate Scale Quantum (NISQ) devices~\cite{cerezo2020variational}. 

Variational algorithms use low depth quantum circuits as a subroutine in a larger classical optimisation and have been applied broadly, including to binary optimisation problems~\cite{farhi2014quantum,arute2020quantum,zhou2018quantum}, training machine learning models~\cite{havlivcek2019supervised,grant2018hierarchical,schuld2020effect}, and obtaining energy spectra~\cite{kandala_hardware-efficient_2017,higgott_variational_2019,ollitrault_quantum_2019}. While low depth circuits lessen the effect of errors, error rates are still a challenge for current practical implementations. Addressing these errors has been a focus within the literature, with many sophisticated error mitigation approaches being developed~\cite{kandala2019error,bravyi2020mitigating,cai2020mitigating}. However, errors will not be the only limiting factor for VQA's. Other obstacles include ansatz construction~\cite{grimsley_adapt_vqe_2019,tang_qubit-adapt-vqe_2020}, optimisation challenges \cite{mcclean_barren_2018,cerezo_cost-function-dependent_2020} and integrated hardware design~\cite{cruise2020practical}.

VQA's are very demanding of quantum hardware, requiring large numbers of sequential calls to quantum devices. In many cases demands on device throughput are further exacerbated by a need to solve multiple different but related optimisation problems (e.g. molecules with different nuclear separations~\cite{kandala_hardware-efficient_2017}, or edge weighted graphs for different weights~\cite{zhou2018quantum}), which to date have mostly been treated independently. Few notable exceptions have solved related problems {iteratively} \cite{zhang2020collective}. Parallelising related problems offers a way to maximize the utility of each circuit. This is particularly relevant to cloud-based interfaces --- rapidly becoming the industry standard --- however, even with dedicated device access, the low-throughput of quantum hardware remains a limiting factor, worsening for larger and more complex problems.


In this paper we introduce a\tcr{\st{ new}} parallel optimisation scheme for VQA problems where the cost function is parameterised by some physical parameter(s). A collection of problems, corresponding to different values of physical parameters, are optimised in parallel by an array of Bayesian optimisers sharing quantum measurement results between them. Using our Bayesian Optimisation with Information Sharing (BOIS) approach, we demonstrate a significant reduction in the number of circuits required to obtain a good solution at all parameter points. BOIS allows us to efficiently find potential energy curves and surfaces for small molecules and a quantum spin model on IBM Quantum devices~\cite{ibmq-devices}.    

\begin{figure}[t]
    \centering
    \includegraphics[width=\columnwidth]{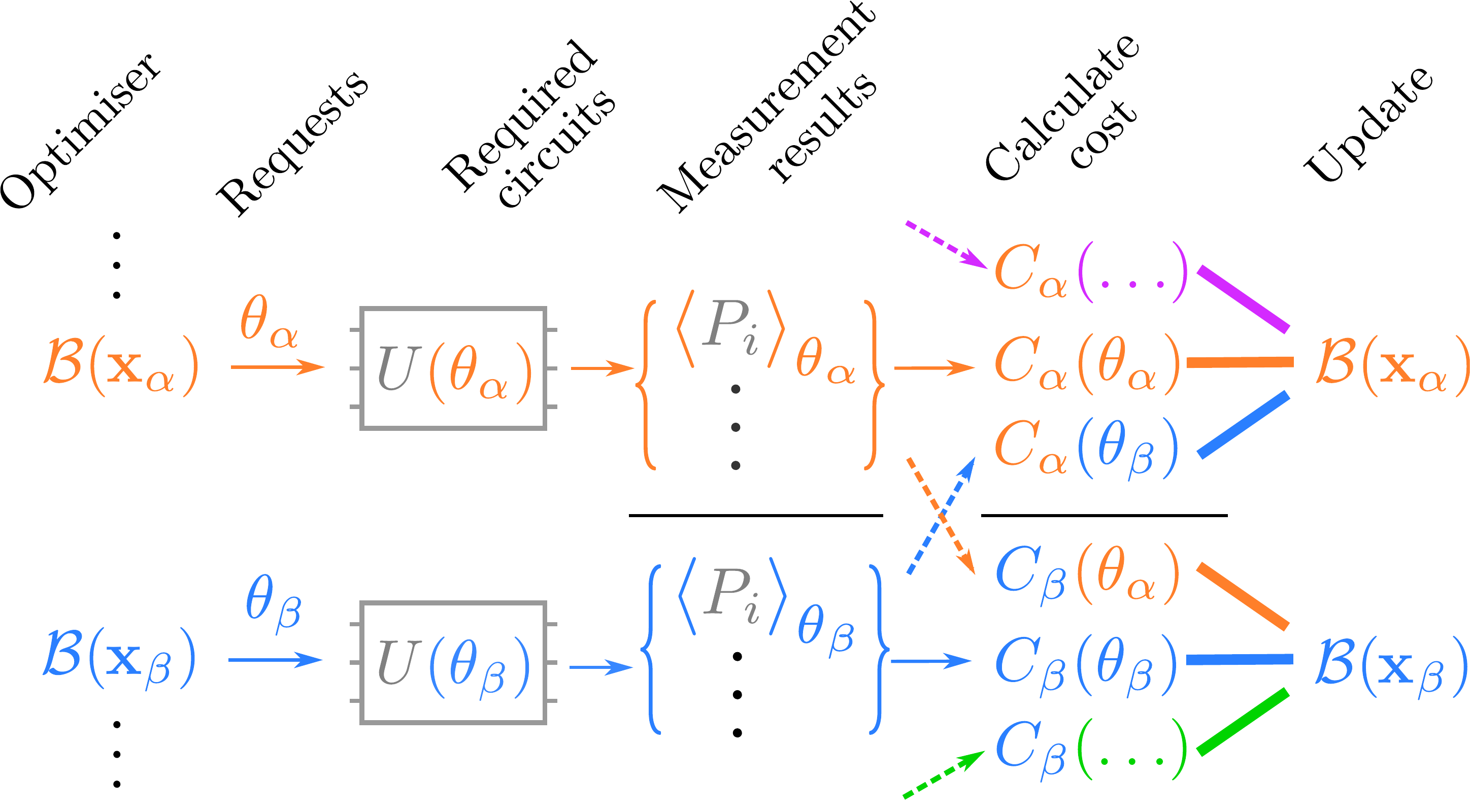}
    \caption{\tcr{Information sharing framework.} Parallel Bayesian optimisation for physically parameterised VQE tasks. Separate BO's $\mathcal{B}(\tcr{\boldsymbol{\mathrm{x}}}_\alpha)$ optimise for different cost functions $C_\alpha$, corresponding to different values of the physical parameters $\tcr{\boldsymbol{\mathrm{x}}}_\alpha$, using the same parameterised ansatz circuit $U(\tcr{\boldsymbol{\theta}})$. Every iteration, each $\mathcal{B}(\tcr{\boldsymbol{\mathrm{x}}}_\alpha)$ requests a new variational parameter point $\tcr{\boldsymbol{\theta}}_\alpha$, at which the set of Pauli strings $\{P_i\}$ are measured.  These expectation values are used to compute any $C_\beta$ cost functions (dashed lines) at $\tcr{\boldsymbol{\theta}}_\alpha$, for all $\alpha,\beta$. Each BO can then be updated using the measurement results obtained for several $\tcr{\boldsymbol{\theta}}_{\alpha},~\tcr{\boldsymbol{\theta}}_{\beta},~\ldots$ parameter points each iteration (bold arrows), dramatically speeding up convergence at all $\tcr{\boldsymbol{\mathrm{x}}}_\alpha$.}
    \label{fig:parallel_BO}
\end{figure}

\begin{figure*}[t]
    \centering
    \includegraphics[width=\textwidth]{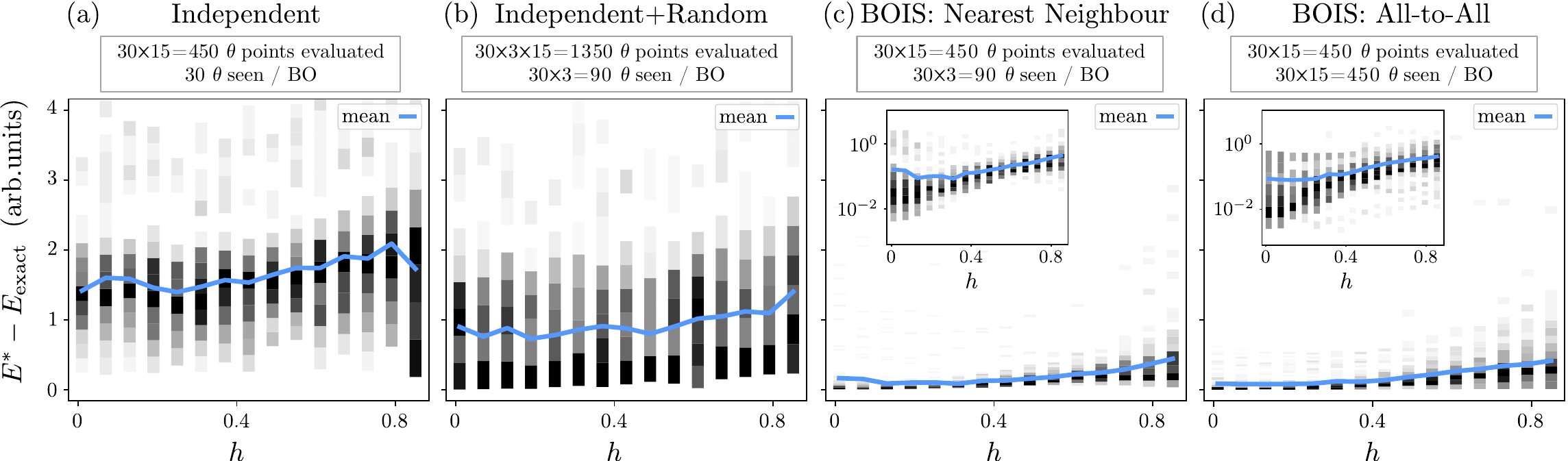}
    \caption{\tcr{Numerical study of the effectiveness of information sharing.} Comparison between (a-b) independent vs (c-d) information shared VQE optimisation strategies for the quantum spin model, Eq.~\eqref{eq:Hising} with field $h=h_X=h_Z$. The sample distribution of 100 repetitions of the same optimisation are shown as density plots for each $h_\alpha$, with darker shades corresponding to more observations, and the solid curve showing the mean. Optimisation is run for thirty iterations and we consider fifteen values of $h_\alpha$. Data boxes show the total number of function evaluations per repetition and the number of evaluations seen by each optimiser $\mathcal{B}(h_\alpha)$, in addition to the numbers quoted each BO receives ten initialisation data points that are also either independent or shared (see methods ~\ref{sec:methods-bo}). (a) Independent BO's at each $h_\alpha$. (b) As before, but each optimiser requests two additional energy evaluations (at random $\tcr{\boldsymbol{\theta}}$ parameter points) at each iteration. (c) BOIS nearest-neighbour information sharing, i.e. BO's at neighbouring $h$-field points (e.g $h_\alpha$ and $h_{\alpha\pm 1}$) share device measurement results. (d) BOIS all-to-all information sharing, i.e. every BO sees all device measurement results. Insets of (c) and (d) show the same data on a log-scale. }
    \label{fig:info-mode-comparison}
\end{figure*}

\section{Results}


\subsection{Parallelising VQE}
\label{sec:par_vqe}


As an example VQA, we consider the Variational Quantum Eigensolver (VQE)~\cite{peruzzo_variational_2014}, used to find the ground states of qubit Hamiltonians $H$. VQE uses a quantum ansatz circuit, parameterised by a set of angles \tcr{$\boldsymbol{\theta}$}, to prepare a state $|\psi(\boldsymbol{\tcr{\theta}})\rangle$, and estimate the expectation value $\mean{H}_{\tcr{\boldsymbol{\theta}}} \equiv\langle \psi(\tcr{\boldsymbol{\theta}}) |H|\psi(\tcr{\boldsymbol{\theta}})\rangle$. Classical optimisation is used to minimize the cost function $\mean{H}_{\tcr{\boldsymbol{\theta}}}$ with respect to $\tcr{\boldsymbol{\theta}}$, which --- for a sufficiently expressive ansatz (see methods~\ref{methods_ansatz}) --- corresponds to the ground state.

Qubit Hamiltonians can always be decomposed into a weighted sum of Pauli strings $P_i \in \{ I,X,Y,Z \}^{\otimes n}$ (for $n$ qubits) with the weights $c_i$ depending on some physical parameter(s) $\tcr{\boldsymbol{\mathrm{x}}}$,
\begin{equation}
H(\boldsymbol{\mathrm{x}}) = \sum_i c_i(\boldsymbol{\mathrm{x}}) P_i.
\label{eq:ham1}
\end{equation}
These physical parameters are, for example, nuclear coordinates of a molecule, length scale of long range interactions, distance of reactants above a catalyst, etc. 

From the structure of Eq.~\eqref{eq:ham1}, it is clear that optimising for each $\tcr{\boldsymbol{\mathrm{x}}}$ independently does not exploit the relationships between problems at different physical parameters. In particular, measuring the required Pauli expectation values $\mean{P_i}_{\tcr{\boldsymbol{\theta}}}$ at a single $\tcr{\boldsymbol{\theta}}$ point can be used to estimate $\langle H(\tcr{\boldsymbol{\mathrm{x}}})\rangle_{\tcr{\boldsymbol{\theta}}}$ for all $x$, as it is simply a weighted sum of Pauli expectation values with weights $c_i(\tcr{\boldsymbol{\mathrm{x}}})$



\subsection{Bayesian optimisation and BOIS}

In order to exploit the relationships between VQE problems, we use an optimisation method that collates function evaluations across the entire $\tcr{\boldsymbol{\theta}}$ parameter space. Bayesian Optimisation (BO) uses a surrogate model of the global cost function to guide the optimization~\cite{frazier_tutorial_2018}. The surrogate model at unevaluated $\tcr{\boldsymbol{\theta}}$ points is inferred, by means of Bayes rules, based on all previous cost functions evaluations. Every optimization round, the next $\tcr{\boldsymbol{\theta}}$ point is decided by maximizing a utility function over the surrogate model (see methods~\ref{sec:methods-bo}). BO has been applied in quantum control~\cite{wigley2016fast,nakamura2019non,mukherjee_bayesian_2020,sauvage_optimal_2020}, as well as in experimental variational algorithms~\cite{otterbach2017unsupervised,zhu2019training}, however its use in VQE and the exploitation of its global surrogate model remains largely unexplored.

BOIS employs an array of Bayesian optimisers running in parallel. Each optimiser, $\mathcal{B}(\tcr{\boldsymbol{\mathrm{x}}}_\alpha)$, attempts to solve a separate but related VQE problem, by minimising $\langle H(\tcr{\boldsymbol{\mathrm{x}}}_\alpha) \rangle_{\tcr{\boldsymbol{\theta}}}$ over $\tcr{\boldsymbol{\theta}}$ at physical coordinate $\tcr{\boldsymbol{\mathrm{x}}}_\alpha$, see Figure~\ref{fig:parallel_BO}. 
At each iteration, optimiser $\mathcal{B}(\tcr{\boldsymbol{\mathrm{x}}}_\alpha)$ requests a cost function evaluation at the Bayes optimal point to evaluate next $\tcr{\boldsymbol{\theta}}_\alpha$ (see methods~\ref{sec:methods-bo}). As described in the section~\ref{sec:par_vqe}, the Pauli string measurement results required to evaluate $\langle \psi(\tcr{\boldsymbol{\theta}}_\alpha) | H(\tcr{\boldsymbol{\mathrm{x}}}_\alpha) | \psi(\tcr{\boldsymbol{\theta}}_\alpha) \rangle$ also allows us to evaluate $\langle \psi(\tcr{\boldsymbol{\theta}}_\alpha) | H(\tcr{\boldsymbol{\mathrm{x}}}_\beta) | \psi(\tcr{\boldsymbol{\theta}}_\alpha) \rangle$, which can be passed to the optimiser $\mathcal{B}(\tcr{\boldsymbol{\mathrm{x}}}_\beta)$. We refer to these cross-evaluations as information sharing from $\mathcal{B}(\tcr{\boldsymbol{\mathrm{x}}}_\alpha)$ to $\mathcal{B}(\tcr{\boldsymbol{\mathrm{x}}}_\beta)$. This requires each optimiser to use the same $\tcr{\boldsymbol{\theta}}$-parameterised ansatz circuit and that $H(\tcr{\boldsymbol{\mathrm{x}}}_\alpha)$ decomposes into the same set of Pauli strings $\{P_i\}$ for all $\alpha$ (this can be padded as necessary).

The intrinsic locality of gradient-based approaches limits the amount of information which can be shared between different optimization runs. While we focused on BO, other surrogate model-based approaches~\cite{sung2020using} could be equivalently lifted to benefit from this information sharing scheme.

\subsection{Testing the effectiveness of information sharing}
\label{sec:info-sharing-effectiveness}

Firstly we demonstrate the effectiveness of information sharing by using BOIS for a VQE task (VQE+BOIS) applied to a quantum spin chain with Hamiltonian
\begin{equation}
    H(h_X,h_Z) = \sum_{<i,j>} Z_i Z_j -  \sum_i (h_X X_i + h_Z Z_i) \, .
    \label{eq:Hising}
\end{equation}
The (dimensionless) transverse $h_X$ and longitudinal $h_Z$ fields are classical coordinates analogous to nuclear separation in molecules. For $h_X>0$ and $h_Z=0$, Eq.~\eqref{eq:Hising} can be diagonalized by a Jordan-Wigner transformation, which transforms the spins into non-interacting fermions. For $h_X,h_Z>0$ it becomes non-integrable, with approximate methods eventually breaking down around a phase transition occurring along a critical line~\cite{ovchinnikov2003antiferromagnetic}.

We consider the case where $h_X = h_Z = h \in [0, 0.9]$, which we discretise into 15 values $h_\alpha$. This range of $h$ approaches but does not cross the critical line. The VQE is then run using our BOIS optimisation approach for a system of four spins, with open boundary conditions. We use an ansatz circuit systematically generated --- using an original approach (see methods~\ref{sec:methods-bo}) --- for this Hamiltonian. Noiseless simulations are carried out by contracting tensor network representations of the ansatz circuit and Hamiltonian using the Quimb python package~\cite{gray_quimb_2018}.

The performance of four different information sharing strategies is shown in Fig.~\ref{fig:info-mode-comparison}. Firstly, as a baseline, we consider independent BO with no information sharing Fig.~\ref{fig:info-mode-comparison}~(a), followed by independent BO's with two extra (random) function evaluations per iteration Fig.~\ref{fig:info-mode-comparison}~(b). This second case is chosen such that the number of evaluations seen by each $\mathcal{B}(h_\alpha)$ is the same as nearest-neighbour information sharing (each $h_\alpha$ shares only with $h_{\alpha \pm 1}$) Fig.~\ref{fig:info-mode-comparison}~(c). Finally we show all-to-all information sharing (each $h_\alpha$ shares with all $h_\beta$, $\beta \neq \alpha$) Fig.~\ref{fig:info-mode-comparison}~(d), which requires the same number of function evaluations as nearest-neighbour sharing but employs more cross-evaluations. Each strategy is run for thirty iterations, after which we compare the final VQE energy estimate ($E^*$) to the exact ground state energy ($E_\textrm{exact}$). Optimisation of the full energy curve is repeated 100 times for each strategy. We present the data as a shaded histogram of $E^* - E_\textrm{exact}$ at each $h_\alpha$, as well as the sample mean. {Each optimisation run is initialised with ten cost function evaluations, these are either shared for the BOIS strategies Fig.~\ref{fig:info-mode-comparison}~(c-d) or independent Fig.~\ref{fig:info-mode-comparison}~(a-b), as discussed in methods~\ref{sec:methods-bo}.}

We see clear improvement of sharing strategies over independent optimisations. After thirty iterations none of the independent BO Fig.~\ref{fig:info-mode-comparison}(a) cases have converged close to the exact ground state energy. Adding extra energy evaluations (at random $\tcr{\boldsymbol{\theta}}$) to the independent optimisations improves their performance Fig.~\ref{fig:info-mode-comparison}(b), highlighting the capacity of BO to leverage the full dataset of evaluations it can access. Despite the improvement, a significant portion of runs end far from the ground state, with non-zero mean error and large variance. Nearest-neighbour sharing Fig.~\ref{fig:info-mode-comparison}(c) hugely improves on Fig.~\ref{fig:info-mode-comparison}(b) despite the individual BO's of the two strategies receiving the same number of function evaluations per iteration. Here, the extra evaluations have greater utility compared to Fig.~\ref{fig:info-mode-comparison}(b) as they come from BO's solving similar (i.e. for close values of $h$) optimisation problems. 
Finally, All-to-All sharing shows moderate improvement over nearest-neighbour sharing, Fig.~\ref{fig:info-mode-comparison}(d) in spite of having received five times more energy evaluations. This indicates that most of the advantage of the information sharing is coming from BO's that are close together in $h$. An additional advantage of nearest-neighbour sharing is it reduces the computational overhead of updating the surrogate model by only including the most relevant shared data. As expected, we find the hardest optimisation problems, where our sharing strategies still do not always converge, to be large values of $h$ closest to the critical line. 

\begin{figure*}
    \centering
    \includegraphics[width=\textwidth]{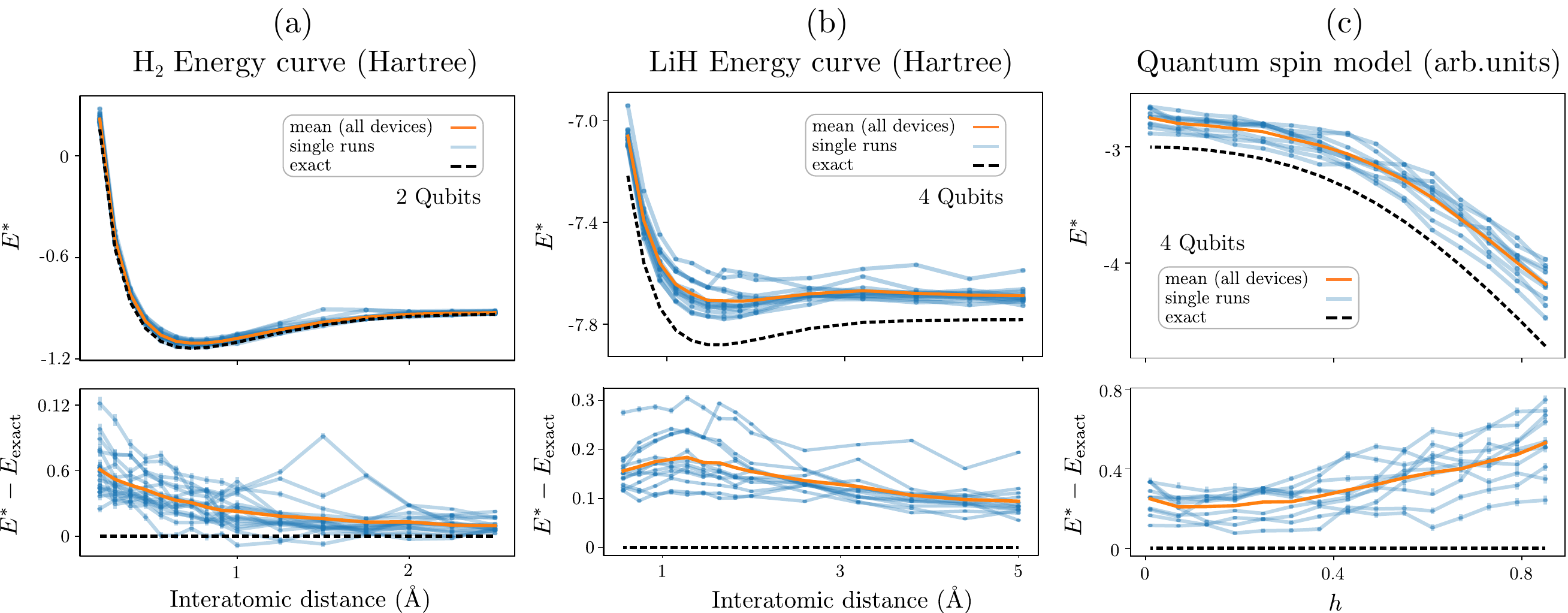}
    \caption{\tcr{Experimental application of VQE+BOIS running on IBMQ devices.} Final energy estimates $E^*$ (top) and final errors $(E^* - E_\textrm{exact})$ (bottom) are plotted for three different problems. Repeated runs (blue curves), executed on a set of different devices~\cite{ibmq-devices}, show the distribution around the mean (orange curves).
    (a) H$_2$ dimer showing eighteen separate VQE+BOIS (all-to-all) runs. Each ran for ten iterations with thirty (shared -- see methods~\ref{sec:methods-bo}) initial points. 
    (b) LiH dimer showing thirteen separate VQE+BOIS (all-to-all) runs. Each ran for thirty iterations with thirty (shared) initial points.
    (c) Quantum spin model, Eqn.~\eqref{eq:ham1} with $h_X = h_Z = h$, (considered in section~\ref{sec:info-sharing-effectiveness})  showing ten separate VQE+BOIS (nearest-neighbour) runs. Each ran for fifty iterations with ten (shared) initial points. Error bars indicate standard error on final energy measurements.}
    \label{fig:1d-experimental-data}
\end{figure*}

\subsection{Experimental demonstration on IBMQ devices}

In the following we demonstrate the success of the BOIS strategy on problems run on IBMQ devices. By studying a quantum spin model (used in section~\ref{sec:info-sharing-effectiveness}) as well as H$_2$, LiH and a linear chain of H$_3$, we highlight the capability of our approach to complete VQE tasks in practical timescales. Optimisation tasks were carried out either on \tcr{Paris, Toronto, Athens, Manhattan, Valencia, or Santiago} IBMQ quantum processors~\cite{ibmq-devices}. We employ Qiskit’s~\cite{ibmq-qiskit} built in `Complete Measurement Filter' with no further error mitigation. {During optimisation 1024 measurement shots are used, increasing to 8192 shots for the final reported VQE energies.}

Figure~\ref{fig:1d-experimental-data} shows the BOIS methods applied to finding ground states of H$_2$ (all-to-all sharing), LiH (all-to-all sharing) and the spin model (nearest-neighbour sharing), each of which only have one physical parameter. These are two, four and four qubit problems respectively, with six, thirteen, and ten optimisation parameters respectively (see methods~\ref{sec:methods-bo}). In obtaining the Hamiltonians for H$_2$ and LiH, we have removed the spin $\mathds{Z}_2$ symmetries, and in the case of LiH, have also frozen non-participating orbitals (see~\ref{sec:molecule_methods}). H$_2$ can be reduced to a single qubit \cite{Bravyi2017Jan}, however for comparison with previous work, the two qubit Hamiltonian is used. These optimisations converge after a remarkable $\sim 10-50$ iterations, far below what could be reasonably expected for any stochastic gradient descent, which would require many hundreds of iterations. Such a drastic improvement unequivocally demonstrates the advantage of data sharing between optimizers, showing just how much information can be extracted from each set of quantum measurements. 

\begin{figure*}[t]
    \centering
    \includegraphics[width=0.96\textwidth]{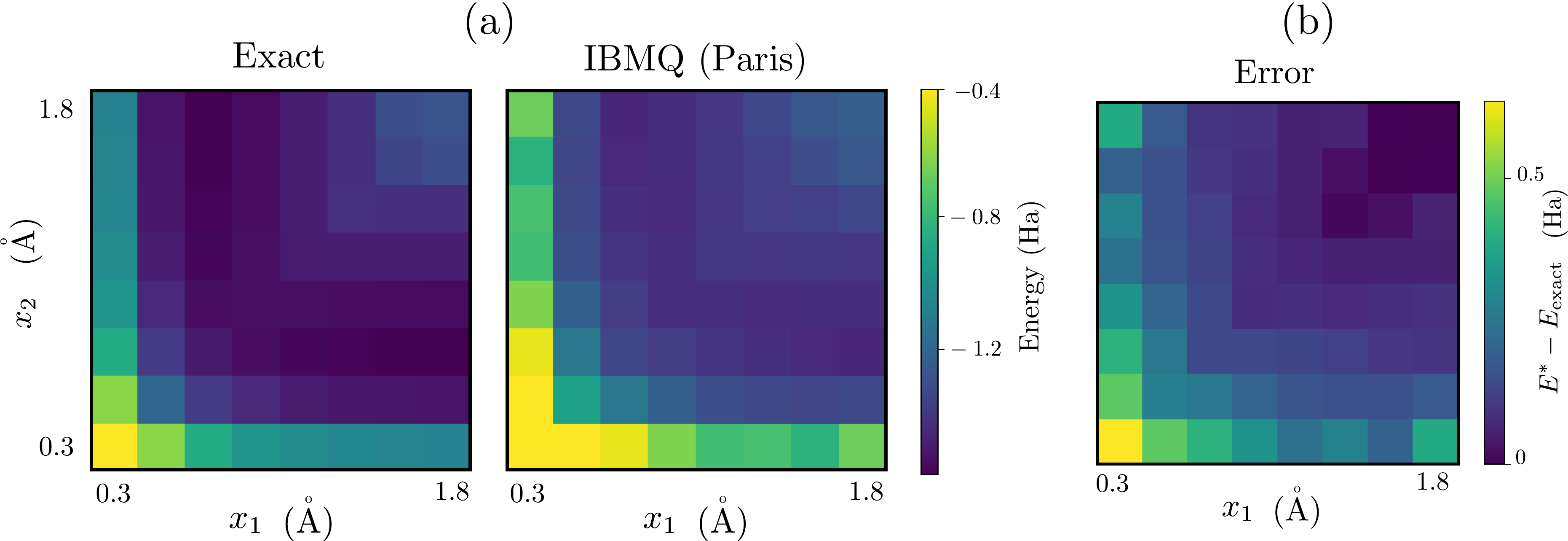}
    \caption{\tcr{Experimental application to a multi-dimensional problem.} Ground state energies (in Hartree) for a linear chain of H$_3$, over two-dimensional parameter space $(x_1,x_2)$ where $x_1$ the distance between the first \& second H atoms and $x_2$ the distance between the second \& third. \tcr{\st{(Left)}(a)} the exact ground state energy surface for an 8$\times$8 grid of $x_1$ and $x_2$ values, compared to experimental results from running VQE+BOIS on from IBM's Paris chip. The optimisations converged in ten iterations to compute the entire surface. \tcr{\st{(Right)}(b)} the error in the VQE+BOIS results, showing systematic offsets due to gate errors.}
    \label{fig:h3-experimental-data}
\end{figure*}

We now consider a linear chain of H$_3$, parameterized by two relative inter-atomic distances $\tcr{\boldsymbol{\mathrm{x}}} = (x_1, x_2)$, which we discretize into a grid to find the 2D energy surface.   Figure~\ref{fig:h3-experimental-data} shows the two-dimensional potential energy surface found using VQE+BOIS with nearest-neighbour sharing. For this two-dimensional problem nearest-neighbour sharing means $(x_1^\alpha,x_2^\beta)$ shares with $(x_1^{\alpha+1},x_2^\beta)$, $(x_1^{\alpha-1},x_2^\beta)$, $(x_1^\alpha,x_2^{\beta+1})$ and $(x_1^\alpha,x_2^{\beta-1})$.
Solving for the eight-by-eight grid in parallel means sixty-four $\tcr{\boldsymbol{\theta}}$ points per iteration, however, we construct the entire surface with just ten iterations, for a total of 650 individual $\tcr{\boldsymbol{\theta}}$ evaluations (including initialisation). In contrast, we expect gradient descent (in this ten dimensional optimisations space) would require a similar number of $\tcr{\boldsymbol{\theta}}$ evaluations just for a single physical coordinate~\cite{kandala_hardware-efficient_2017}. 


\section{Discussion}

Our experimental results, Fig.~\ref{fig:1d-experimental-data}, show a discrepancy between experimental and theoretical energy curves/surfaces. The most significant sources of VQE errors are (i) optimisation errors, and (ii) device errors. Ansatz expressibility errors, the other limiting factor for VQE, can be ruled out by our custom circuit design (see methods~\ref{methods_ansatz}). Known difficulties in the optimisation landscapes mean single runs of the optimisation will sometimes fail~\cite{mcclean_barren_2018}. This is exemplified in Fig.~\ref{fig:info-mode-comparison} for large $h$ where the ground state is more entangled. The same qualitative behaviour is observed in the experimental data Fig.~\ref{fig:1d-experimental-data} (c) and is made worse by the randomness from using a noisy quantum device to estimate the cost function.

The systematic offset in the experimental data, Fig.~\ref{fig:1d-experimental-data}, is typical of depolarising errors in quantum gates, becoming worse for more qubits~\cite{kandala2019error}. Variations between runs are expected as different devices have different error characteristics and additionally these change over time. In general, the variance is further compounded by shot noise (i.e. finite numbers of measurement on device), but this is well below the variance of device noise across different days and devices.

Why would we expect this information sharing to be helpful? Considering a molecule as an example, a small change in the nuclear coordinates can be understood as a perturbation of the electronic Hamiltonian. The perturbed and unperturbed ground states will have some overlap, so their optimal $\tcr{\boldsymbol{\theta}}$'s will be found in a similar region of parameter space. We can then see that a useful $\tcr{\boldsymbol{\theta}}_\alpha$ selected by optimiser $\mathcal{B}(\tcr{\boldsymbol{\mathrm{x}}}_\alpha)$ is likely to be valuable to $\mathcal{B}(\tcr{\boldsymbol{\mathrm{x}}}_\alpha+\delta \tcr{\boldsymbol{\mathrm{x}}})$. In this way, information sharing allows each BO to exploit the promising regions of $\tcr{\boldsymbol{\theta}}$-space discovered by other optimisers, making the evaluations of adjacent optimisers more valuable than random $\tcr{\boldsymbol{\theta}}$ evaluations.

Information sharing would be particularly helpful when the physical parameter tunes some non-trivial interaction that continuously increases the entanglement in the ground state, for example approaching a phase transition as in section~\ref{sec:info-sharing-effectiveness}. In this case we would expect the optimisation to converge faster in the simple (less entangled ground state) limit and this information to flow to harder (more entangled) problems. 

Here we introduced BOIS, an efficient \tcr{\st{new }}optimisation scheme for related VQA's. Our approach is specifically designed to make maximum use of limited numbers of quantum measurement in order to combat the limited throughput of NISQ quantum devices. We demonstrated the efficiency of our approach by experimentally solving for energy curves in tens, instead of hundreds or thousands of iterations. The speedup provided by our method has enabled both 1D energy curves and the 2D energy surface of a molecular trimer to be found on practical timescale, even without dedicated device access. BOIS can be readily applied to other parameterised VQA's and, combined with our circuit construction, makes benchmarking VQA algorithms on real devices significantly more practical.


\section{Methods}

\subsection{Qubit Hamiltonians for small molecules}
\label{sec:molecule_methods}

To find molecular ground states we first reformulate the problem from electrons to qubits. This is discussed at length in the literature, e.g. in the reviews~\cite{mcardle2020quantum,bauer2020quantum}. Here we outline the general procedure and highlight features relevant to the above simulations. 

The electronic states are expanded in hydrogen-like orbitals centered on each atom, specifying a fermionic operator and a spatial wave function for each spin orbital. The electronic spatial dependence can be integrated out under the Born-Oppenheimer approximation. This re-expresses the kinetic and potential energy terms as non-linear coupling rates between the fermionic operators, where the couplings rates are nuclear coordinates dependent. The Hilbert space can be further reduced by removing or freezing (i.e. permanently occupied) spin orbitals. Finally, the anti-commuting fermionic operators are transformed into commuting Pauli operators using the Jordan-Wigner, Bravi-Kitaev, or some similar map~\cite{mcardle2020quantum}.

In this work we use Openfermion~\cite{openfermion} (for H$_3$) and Qiskit~\cite{ibmq-qiskit} (for H$_2$ and LiH) to compute the one and two body integrals, using the STO-3G approximation. For H$_2$ and H$_3$, only the 1s orbitals are used (none of which are frozen), and the systems have a total of two and three electrons respectively. For LiH, the Li:1s, orbital is doubly occupied with electrons and frozen, while the four Li:2p$_y$, Li:2p$_z$ spin orbitals are forzen with zero occupation. Thus only the four Li:2s,2p$_x$ spin orbitals (bond axes aligned with $x$) and two H:1s spin orbitals constitute the available states for the remaining two electrons. We use the symmetry conserving Bravi-Kitaev transform for H$_3$, and the parity transform for H$_2$ and LiH \tcr{to express the fermionic Hamiltonian as a qubit Hamiltonian}. For each molecule the $\mathds{Z}_2$ symmetries (corresponding to particle and spin symmetries) have been removed~\cite{kandala_hardware-efficient_2017,Bravyi2017Jan}, reducing the number of qubits by two.

\subsection{Ansatz design}
\label{methods_ansatz}

\begin{figure*}[t]
    \centering
    \includegraphics[width=\textwidth]{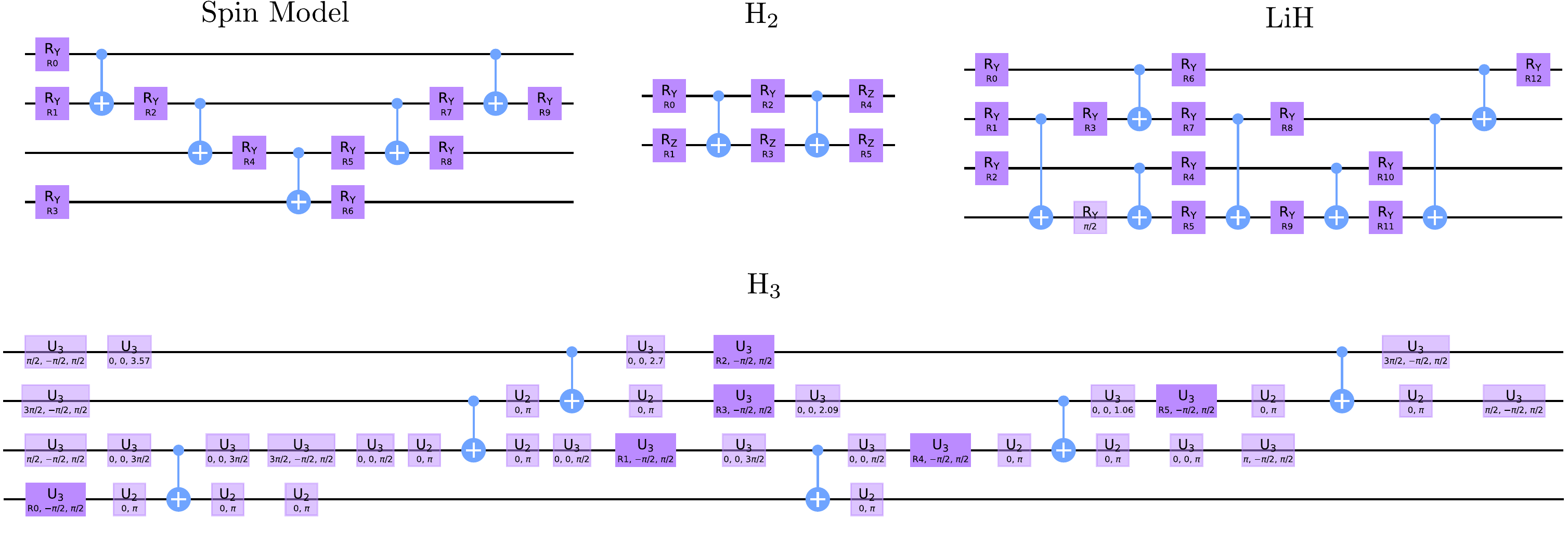}
    \caption{\tcr{Parameterised quantum circuits used as ansatz.} Ansatz circuits used for each of the physical systems we consider, generated using the procedure described in section~\ref{methods_ansatz}. Gates referenced are from the Qiskit gate set~\cite{ibmq-qiskit}. Parameterised gates are shown in solid colours and fixed angle rotations are indicated with partial transparency.}
    \label{fig:ansatz}
\end{figure*}

As our focus is on efficient optimisation, rather than the challenging and distinct problem of blind ansatz construction, we construct ansatzes that are tailored to the problems we consider. The effectiveness of VQE is highly dependent on the suitability of the ansatz used. An ansatz circuit must be sufficiently expressive to produce its target state, however this must be balanced against other practical concerns including hardware limitations (restricted qubit connectivity and practical limits on circuit depth due to experimental noise) as well as optimisation considerations (the number of optimisation parameters must not be too large). 

Our ansatzes, for systems other than H$_2$, are constructed classically such that they are guaranteed to be able to produce the target states with high fidelity (in noiseless simulations), while being both depth and parameter number efficient. This allows us to assess the performance of our optimisation scheme without it being limited by the expressive power of the ansatz. Additionally, this allows us to bake hardware constraints, such as qubit connectivity, into the construction process. The ansatzes we use for each of the systems we consider are shown in Fig.~\ref{fig:ansatz}.

Building the ansatz circuit is done in two phases; a growth phase in which the expressivity of the circuit is increased, and a shrinkage phase in which redundant parameters are removed. This approach is inspired by qubit-ADAPT \cite{tang_qubit-adapt-vqe_2020} but uses exhaustive optimisation to decide how to grow the circuit rather than a gradient-based condition. The noiseless simulations used to construct these ansatzes are carried out with highly efficient tensor-network calculations~\cite{bridgeman_tensor-networks_2017}, performing gradient-descent with the Quimb Python package~\cite{gray_quimb_2018}. We now describe each phase in detail. 

The growth phase begins with a separable circuit consisting only of single-qubit rotations and adds two-qubit entangling blocks in the locations that provide the greatest increase in the fidelity with the target state. Entangling blocks consist of a two-qubit gate, e.g. CNOT, padded on either side with parameterised single-qubit rotations. In general, the single-qubit gates are arbitrary Bloch sphere rotations (e.g. Qiskit's $U3$ gate or equivalent) to allow maximum flexibility. However, in many cases simpler gates can be used (such as real-valued Hamiltonians where $R_Y$ is sufficient).
At each step, we determine all locations that it is possible to place a new entangler, compatible with the qubit connectivity. Early in the optimisation this is a weak constraint as we do not initially enforce a mapping of virtual to physical qubits. As the ansatz grows the mapping is fixed by the previously chosen locations and this greatly reduces the number of potential locations in subsequent steps. We test each possible location by optimising the fidelity $\mathcal{F}$ between the target state and the circuit with this block added. For small to moderately sized systems, noiseless tensor-network-based gradient descent can be performed efficiently with automatic differentiation (e.g. using Google's TensorFlow~\cite{tensorflow_2015}) to minimise a cost $\mathcal{C}$. For a target state $\ket{\psi_T}$ and trial state \ket{\psi} we used
\begin{equation}
C=1-\mathcal{F}=1-|\langle\psi|\psi_t\rangle|.
\end{equation}
(Here we use this convention for the fidelity, as opposed to $|\langle\psi|\psi_t\rangle|^2$, as this yielded better performance.) The entangling block that leads to the greatest increase in fidelity is added to the ansatz and the process is repeated. Since our scheme relies on direct optimisation we are able to trial placing entangling blocks at both the beginning and end of the circuit. Placing an entangling block at the beginning of a circuit can produce a more dramatic change than placing one at the end, potentially allowing our method to converge to the target state more quickly than other greedy methods such as qubit-ADAPT. The ansatz that is produced in the growth phase is very efficient in terms of depth. Using gradient descent the growth stage can be performed with relatively few cost/gradient evaluations.

Once the circuit has converged to an acceptable fidelity with the target state (we used $1-\mathcal{F}<10^{-6}$) we begin the shrinkage phase. We re-optimise the fidelity of the ansatz with respect to the target state but now adding a regularisation penalty to our cost function which, for gate angles $\{\phi_i\}$, becomes
\begin{equation}
C=1-\mathcal{F}+\eta\sum_i D(\phi_i, 2\pi)
\end{equation} 
Here $\eta$ is a small regularisation weight parameter and $D(\phi_i, 2\pi)$ is the absolute distance between $\phi_i$ and its nearest multiple of $2\pi$. This is an example of (periodic) $L_1$ regularisation and, provided $\eta$ is small, encourages gate angles to shrink towards a multiple of $2\pi$ while maintaining a high fidelity. This allows us to identify unnecessary single-qubit gates (those with angles close to multiples of $2\pi$) which are then removed from the ansatz. This regularisation process is repeated with increasing $\eta$ until the point where removing any more single-qubit gates would appreciably degrade the fidelity (checking if this is the case by re-optimising the fidelity using the ansatz with small parameter gates removes).

The ansatz construction process is non-deterministic however for the systems considered we found that typical required numbers of cost/gradient evaluations were (for the growth/shrinkage stages respectively) -- H$_3$: $\sim\!4500, 1200$ (using general $U3$ gates); LiH $\sim\!1800, 600$ ; quantum spin model $\sim\!1800, 500$.

These ansatzes are produced by optimising the fidelity with a single target state. In our VQE+BOIS simulations we are attempting to find ground state energy surfaces across a physical parameter space. For each VQE+BOIS simulation a target state was chosen as a ground state sitting on the energy surface in question. However, these target states were chosen such that their physical parameter $\tcr{\boldsymbol{\mathrm{x}}}$ was not one that was associated with a BOIS optimiser $\tcr{\boldsymbol{\mathrm{x}}} \notin \tcr{\boldsymbol{\mathrm{x}}}_\alpha$. Although our ansatzes were produced targeting just one state on each energy surface, we found that provided it can represent this target state with a high fidelity (around $1-\mathcal{F}<10^{-6}$) then it will typically have similar performance on the other states of the surface (assuming we do not cross any complicated quantum phase transitions). Finally, to further reduce the number of optimisation parameters, the optimal gate parameters for each BOIS physical parameter $\tcr{\boldsymbol{\mathrm{x}}}_\alpha$ are found (again by maximising the fidelity) and any single-qubit gate angles within this optimal parameter set that are found to remain effectively constant across the energy surface are fixed to these constant values in the ansatz. These fixed angle gates are indicated with partial transparency in Fig.~\ref{fig:ansatz}.

While this ansatz preparation procedure uses the fidelity with a known target state as a cost function, requiring knowledge of the target state, this is not an \tcr{a priori} requirement. With minimal adjustments the same tensor-network-based framework can be adapted to use the expectation of a Hamiltonian as a cost function to be minimised, resulting in a scheme that is qualitatively similar to hardware-efficient ADAPT-VQE \cite{tang_qubit-adapt-vqe_2020}. An energy based cost function would be less specific than one based on the fidelity with a target state and so convergence issues may arise if the Hamiltonian possesses many near-degeneracies. The lifting of degeneracies across physical parameter space may also prevent an energy optimised ansatz from generalising well. Ultimately these classical gradient-based methods are limited to systems of only up to around 10 qubits (we have tested up to this size). 

As we aimed to build tailored VQE ansatzes to benchmark BOIS, rather than develop a full scalable adaptive ansatz algorithm, we employed the more reliable fidelity-based approach. However we believe an adaptive scheme in which an ansatz is grown ansatz one entangling block at a time followed by parameter regularisation has the potential to be run in a more scalable way on-device. Each stage of such a process would be effectively its own VQE problem, potentially allowing information sharing to increase its efficiency, provided that the optimisation is done globally such that information can be shared. By considering all physical parameter points when choosing entangling block placement and performing regularisation, a general ansatz for the whole energy surface may also be obtained more easily than when considering only a single physical parameter point.

\subsection{Practicals aspects of Bayesian Optimisation}
\label{sec:methods-bo}

In our work Bayesian optimisation is implemented using the GPyOpt python package~\cite{gpyopt2016}.
We have used the standard multipurpose Matern-5/2 kernel. Whether more specialised kernels are more suitable for VQA's is an interesting question.

{Bayesian optimisation begins with a pool of initial data. These are cost functions evaluations made at random in the optimisation parameter space, that are used to initialise the surrogate model. We select the $M$ initial points by Latin hypercube sampling. When considering our information sharing strategies we additionally share these $M$ initial points amongst the different optimisers $\mathcal{B}(\tcr{\boldsymbol{\mathrm{x}}}_i)$. In contrast, when we compare this to independent strategies with no information sharing each optimiser $\mathcal{B}(\tcr{\boldsymbol{\mathrm{x}}}_i)$ sees a different set of $M$ initial points.}

In all of our results we use a lower-confidence bound (LCB) acquisition function~\cite{snoek2012practical}, defined by 
\begin{equation*}
    a_\textrm{LCB}(\theta) = \mu(\theta) - \kappa \, \sigma(\theta) \, ,
\end{equation*}
where $\mu$ and $\sigma^2$ are the mean and variance functions of the Gaussian process surrogate model. The hyperparameter $\kappa$ is chosen to decrease linearly from an initial value $\kappa_0$ to zero at the final iteration $N$, so that at iteration $t \in [1,N]$ it has value
\begin{equation*}
    \kappa_t = \kappa_0 \frac{N - t}{N} \, .
\end{equation*}
This highly weights \emph{exploration} early in the optimisation, while prioritising \emph{exploitation} later on. The value of $\kappa_0$ is tuned between one and five depending on the problem. At iteration $t$ the next $\theta$ to evaluate is selected by minimising $a_\textrm{LCB}(\theta)$.

Bayesian optimisation is conventionally limited to small numbers of parameters, typically $< 20-30$~\cite{frazier_tutorial_2018}. This will become a problem for larger VQE problems, especially when using hardware efficient ansatz. 
However, decreasing noise rates will help alleviate this issue, since it will allow more complex but parameter efficient ansatz, such as Unitary Coupled Cluster (UCC)~\cite{mcardle2020quantum}, to be employed. 
Beyond this, global optimisation with large numbers of parameters is an active research topic. Replacing the Gaussian Proccess with other surrogate models, such as neural networks or random forests, can potentially allow much larger numbers of parameters~\cite{shahriari2015taking}.

Problems such as over-parameterisation can, in principle, be solved by using physically motivated ansatz circuits, such as UCC for molecules~\cite{bartlett_alternative_1989}, or schemes that grow an ansatz systematically, for example ADAPT-VQE~\cite{grimsley_adapt_vqe_2019,tang_qubit-adapt-vqe_2020}. 
These approaches typically require circuits that are too deep to be practically run on current NISQ devices, however more sophisticated schemes such as ansatz that preserve symmetries~\cite{yeter2021benchmarking} may help to overcome these challenges.

\begin{figure*}[t]
    \centering
    \includegraphics[width=\textwidth]{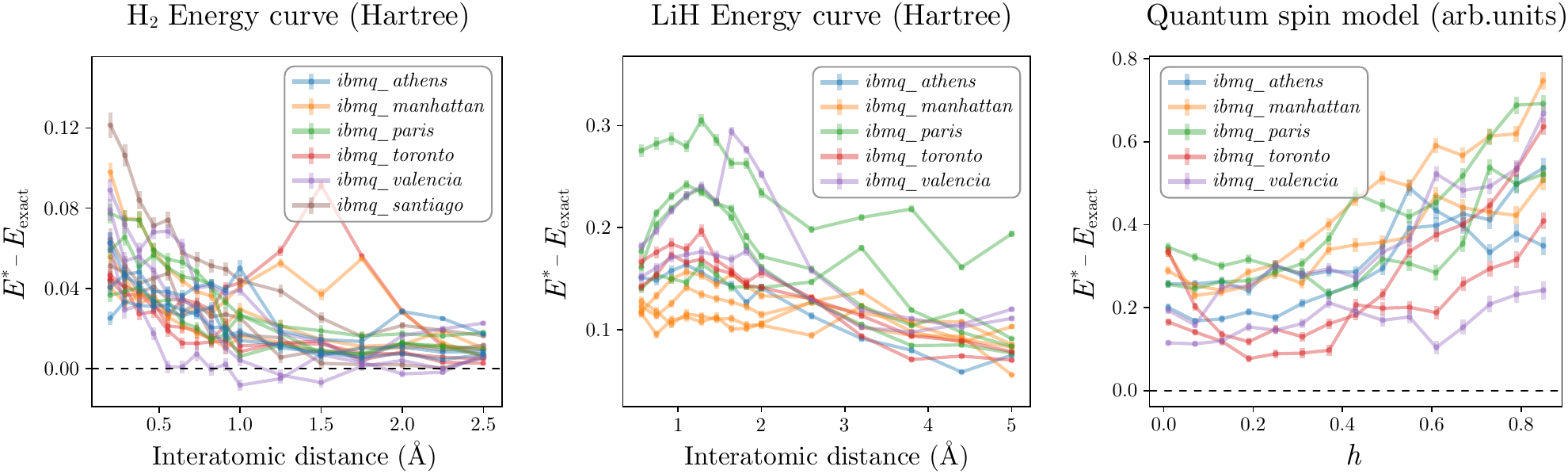}
    \caption{\tcr{Experimental data separated by IBM Quantum device.}
    Final errors, $E^* - E_\textrm{exact}$, in repeated executions of VQE+BOIS ground state energy estimates for H$_2$, LiH and the quantum spin model with the IBM Quantum device used indicated by plot colour. We do not observe any clear, systematic differences between the different devices. Error bars indicate standard error on final energy measurements.
    }
    \label{fig:errors-by-device}
\end{figure*}

\subsection{Separating experimental data by device}

The experimental data presented in Fig.~\ref{fig:1d-experimental-data} was collected in October and November 2020 from the IBM quantum devices Athens, Manhattan, Paris, Toronto and Valencia, with additional H$_2$ data from Santiago~\cite{ibmq-devices}. In Fig.~\ref{fig:errors-by-device} the final VQE errors, $E^* - E_\textrm{exact}$, is plotted for the H$_2$, LiH and quantum spin model tasks, with the IBM device the data was collected from indicated by plot colour.

All of the devices we use have quantum volume 32, with the exception of Valencia which has quantum volume 16. Within our data we do not see clear systematic differences between the devices when looking across all the tasks. For a given task there can appear to be clustering. For example, in our LiH data Paris appears to do consistently worse and Manhattan better. However, this is more likely related to calibration differences in the time window of use or small sample sizes. It is interesting to note, though, that the only non-physical results (ground state energies lower than the true ground state energy) occur for the H$_2$ task on the lower quantum volume device, Valencia.

\section*{Data availability statement}

The data that support the findings of this study are available from the corresponding author upon reasonable request.

\section*{Code availability statement}

The code needed to reproduce the findings of this study are available from the corresponding author upon reasonable request.

\section*{Acknowledgments}
We thank Rick Mukherjee for useful discussions on Bayesian optimisation. 
This work is supported by Samsung GRC grant, the UK Hub in Quantum Computing and Simulation, part of the UK National Quantum Technologies Programme with funding from UKRI EPSRC grant EP/T001062/1 and the QuantERA ERA-NET Cofund in Quantum Technologies implemented within the European Union's Horizon 2020 Programme. 
The project TheoryBlind Quantum Control TheBlinQC has received funding from the QuantERA ERA-NET Cofund in Quantum Technologies implemented within the European Unions Horizon 2020 Programme and from EPSRC under the grant EP/R044082/1.
F.S. is supported by a studentship in the Quantum Systems Engineering Skills and Training Hub at Imperial College London funded by EPSRC (EP/P510257/1).
We acknowledge the use of IBM Quantum services for this work. The views expressed are those of the authors, and do not reflect the official policy or position of IBM or the IBM Quantum team.
Numerical simulations in section~\ref{sec:info-sharing-effectiveness} were carried out on Imperial HPC facilities~\cite{imperial-rcs-acknowledgement}.

\section*{Author Contributions}

All authors contributed to developing the idea of information sharing. Code implementations of Bayesian optimisation with information sharing were written by KK, CS, FS and AS. AS developed and implemented the ansatz design scheme. Simulations and experiments were executed by CS and KK. KK and CS wrote the manuscript. CS, KK and AS wrote the supplementary methods. FM and FS provided expertise on Bayesian optimisation. PH and JK provided theoretical support on the physical problems used as benchmarks. MSK led and coordinated the project. All authors contributed to revising the manuscript.

\section*{Competing Interests}

The authors declare that there are no competing interests.


\begin{thebibliography}{10}
\expandafter\ifx\csname url\endcsname\relax
  \def\url#1{\texttt{#1}}\fi
\expandafter\ifx\csname urlprefix\endcsname\relax\def\urlprefix{URL }\fi
\providecommand{\bibinfo}[2]{#2}
\providecommand{\eprint}[2][]{\url{#2}}

\bibitem{jurcevic2020demonstration}
\bibinfo{author}{Jurcevic, P.} \emph{et~al.}
\newblock \bibinfo{title}{Demonstration of quantum volume 64 on a
  superconducting quantum computing system}.
\newblock \emph{\bibinfo{journal}{arXiv preprint arXiv:2008.08571}}
  (\bibinfo{year}{2020}).

\bibitem{arute2019quantum}
\bibinfo{author}{Arute, F.} \emph{et~al.}
\newblock \bibinfo{title}{Quantum supremacy using a programmable
  superconducting processor}.
\newblock \emph{\bibinfo{journal}{Nature}} \textbf{\bibinfo{volume}{574}},
  \bibinfo{pages}{505--510} (\bibinfo{year}{2019}).

\bibitem{pino2020demonstration}
\bibinfo{author}{Pino, J.} \emph{et~al.}
\newblock \bibinfo{title}{Demonstration of the trapped-ion quantum ccd computer
  architecture}.
\newblock \emph{\bibinfo{journal}{Nature}} \textbf{\bibinfo{volume}{592}},
  \bibinfo{pages}{209--213} (\bibinfo{year}{2021}).

\bibitem{preskill_quantum_2018}
\bibinfo{author}{Preskill, J.}
\newblock \bibinfo{title}{Quantum computing in the nisq era and beyond}.
\newblock \emph{\bibinfo{journal}{Quantum}} \textbf{\bibinfo{volume}{2}},
  \bibinfo{pages}{79} (\bibinfo{year}{2018}).

\bibitem{benedetti2019parameterized}
\bibinfo{author}{Benedetti, M.}, \bibinfo{author}{Lloyd, E.},
  \bibinfo{author}{Sack, S.} \& \bibinfo{author}{Fiorentini, M.}
\newblock \bibinfo{title}{Parameterized quantum circuits as machine learning
  models}.
\newblock \emph{\bibinfo{journal}{Quantum Sci. Technol.}}
  \textbf{\bibinfo{volume}{4}}, \bibinfo{pages}{043001} (\bibinfo{year}{2019}).

\bibitem{smith_simulating_2019}
\bibinfo{author}{Smith, A.}, \bibinfo{author}{Kim, M.},
  \bibinfo{author}{Pollmann, F.} \& \bibinfo{author}{Knolle, J.}
\newblock \bibinfo{title}{Simulating quantum many-body dynamics on a current
  digital quantum computer}.
\newblock \emph{\bibinfo{journal}{npj Quantum Inf.}}
  \textbf{\bibinfo{volume}{5}}, \bibinfo{pages}{1--13} (\bibinfo{year}{2019}).

\bibitem{mcardle2020quantum}
\bibinfo{author}{McArdle, S.}, \bibinfo{author}{Endo, S.},
  \bibinfo{author}{Aspuru-Guzik, A.}, \bibinfo{author}{Benjamin, S.~C.} \&
  \bibinfo{author}{Yuan, X.}
\newblock \bibinfo{title}{Quantum computational chemistry}.
\newblock \emph{\bibinfo{journal}{Rev. Mod. Phys.}}
  \textbf{\bibinfo{volume}{92}}, \bibinfo{pages}{015003}
  (\bibinfo{year}{2020}).

\bibitem{bauer2020quantum}
\bibinfo{author}{Bauer, B.}, \bibinfo{author}{Bravyi, S.},
  \bibinfo{author}{Motta, M.} \& \bibinfo{author}{Kin-Lic~Chan, G.}
\newblock \bibinfo{title}{Quantum algorithms for quantum chemistry and quantum
  materials science}.
\newblock \emph{\bibinfo{journal}{Chem. Rev.}}
  \textbf{\bibinfo{volume}{120}}, \bibinfo{pages}{12685--12717}
  (\bibinfo{year}{2020}).

\bibitem{cerezo2020variational}
\bibinfo{author}{Cerezo, M.} \emph{et~al.}
\newblock \bibinfo{title}{Variational quantum algorithms}.
\newblock \emph{\bibinfo{journal}{arXiv preprint arXiv:2012.09265}}
  (\bibinfo{year}{2020}).

\bibitem{farhi2014quantum}
\bibinfo{author}{Farhi, E.}, \bibinfo{author}{Goldstone, J.} \&
  \bibinfo{author}{Gutmann, S.}
\newblock \bibinfo{title}{A quantum approximate optimization algorithm}.
\newblock \emph{\bibinfo{journal}{arXiv preprint arXiv:1411.4028}}
  (\bibinfo{year}{2014}).

\bibitem{arute2020quantum}
\bibinfo{author}{Harrigan, M.~P.} \emph{et~al.}
\newblock \bibinfo{title}{Quantum approximate optimization of non-planar graph
  problems on a planar superconducting processor}.
\newblock \emph{\bibinfo{journal}{Nat. Phys.}}
  \textbf{\bibinfo{volume}{17}}, \bibinfo{pages}{332--336}
  (\bibinfo{year}{2021}).

\bibitem{zhou2018quantum}
\bibinfo{author}{Zhou, L.}, \bibinfo{author}{Wang, S.-T.},
  \bibinfo{author}{Choi, S.}, \bibinfo{author}{Pichler, H.} \&
  \bibinfo{author}{Lukin, M.~D.}
\newblock \bibinfo{title}{Quantum approximate optimization algorithm:
  Performance, mechanism, and implementation on near-term devices}.
\newblock \emph{\bibinfo{journal}{Phys. Rev. X}}
  \textbf{\bibinfo{volume}{10}}, \bibinfo{pages}{021067}
  (\bibinfo{year}{2020}).

\bibitem{havlivcek2019supervised}
\bibinfo{author}{Havl{\'\i}{\v{c}}ek, V.} \emph{et~al.}
\newblock \bibinfo{title}{Supervised learning with quantum-enhanced feature
  spaces}.
\newblock \emph{\bibinfo{journal}{Nature}} \textbf{\bibinfo{volume}{567}},
  \bibinfo{pages}{209--212} (\bibinfo{year}{2019}).

\bibitem{grant2018hierarchical}
\bibinfo{author}{Grant, E.} \emph{et~al.}
\newblock \bibinfo{title}{Hierarchical quantum classifiers}.
\newblock \emph{\bibinfo{journal}{npj Quantum Inf.}}
  \textbf{\bibinfo{volume}{4}}, \bibinfo{pages}{1--8} (\bibinfo{year}{2018}).

\bibitem{schuld2020effect}
\bibinfo{author}{Schuld, M.}, \bibinfo{author}{Sweke, R.} \&
  \bibinfo{author}{Meyer, J.~J.}
\newblock \bibinfo{title}{Effect of data encoding on the expressive power of
  variational quantum-machine-learning models}.
\newblock \emph{\bibinfo{journal}{Phys. Rev. A}}
  \textbf{\bibinfo{volume}{103}}, \bibinfo{pages}{032430}
  (\bibinfo{year}{2021}).

\bibitem{kandala_hardware-efficient_2017}
\bibinfo{author}{Kandala, A.} \emph{et~al.}
\newblock \bibinfo{title}{Hardware-efficient variational quantum eigensolver
  for small molecules and quantum magnets}.
\newblock \emph{\bibinfo{journal}{Nature}} \textbf{\bibinfo{volume}{549}},
  \bibinfo{pages}{242--246} (\bibinfo{year}{2017}).

\bibitem{higgott_variational_2019}
\bibinfo{author}{Higgott, O.}, \bibinfo{author}{Wang, D.} \&
  \bibinfo{author}{Brierley, S.}
\newblock \bibinfo{title}{Variational quantum computation of excited states}.
\newblock \emph{\bibinfo{journal}{Quantum}} \textbf{\bibinfo{volume}{3}},
  \bibinfo{pages}{156} (\bibinfo{year}{2019}).

\bibitem{ollitrault_quantum_2019}
\bibinfo{author}{Ollitrault, P.~J.} \emph{et~al.}
\newblock \bibinfo{title}{Quantum equation of motion for computing molecular
  excitation energies on a noisy quantum processor}.
\newblock \emph{\bibinfo{journal}{Phys. Rev. Res.}}
  \textbf{\bibinfo{volume}{2}}, \bibinfo{pages}{043140} (\bibinfo{year}{2020}).

\bibitem{kandala2019error}
\bibinfo{author}{Kandala, A.} \emph{et~al.}
\newblock \bibinfo{title}{Error mitigation extends the computational reach of a
  noisy quantum processor}.
\newblock \emph{\bibinfo{journal}{Nature}} \textbf{\bibinfo{volume}{567}},
  \bibinfo{pages}{491--495} (\bibinfo{year}{2019}).

\bibitem{bravyi2020mitigating}
\bibinfo{author}{Bravyi, S.}, \bibinfo{author}{Sheldon, S.},
  \bibinfo{author}{Kandala, A.}, \bibinfo{author}{Mckay, D.~C.} \&
  \bibinfo{author}{Gambetta, J.~M.}
\newblock \bibinfo{title}{Mitigating measurement errors in multiqubit
  experiments}.
\newblock \emph{\bibinfo{journal}{Phys. Rev. A}}
  \textbf{\bibinfo{volume}{103}}, \bibinfo{pages}{042605}
  (\bibinfo{year}{2021}).

\bibitem{cai2020mitigating}
\bibinfo{author}{Cai, Z.}, \bibinfo{author}{Xu, X.} \&
  \bibinfo{author}{Benjamin, S.~C.}
\newblock \bibinfo{title}{Mitigating coherent noise using pauli conjugation}.
\newblock \emph{\bibinfo{journal}{npj Quantum Inf.}}
  \textbf{\bibinfo{volume}{6}}, \bibinfo{pages}{1--9} (\bibinfo{year}{2020}).

\bibitem{grimsley_adapt_vqe_2019}
\bibinfo{author}{Grimsley, H.~R.}, \bibinfo{author}{Economou, S.~E.},
  \bibinfo{author}{Barnes, E.} \& \bibinfo{author}{Mayhall, N.~J.}
\newblock \bibinfo{title}{An adaptive variational algorithm for exact molecular
  simulations on a quantum computer}.
\newblock \emph{\bibinfo{journal}{Nat. Commun.}}
  \textbf{\bibinfo{volume}{10}}, \bibinfo{pages}{1--9} (\bibinfo{year}{2019}).

\bibitem{tang_qubit-adapt-vqe_2020}
\bibinfo{author}{Tang, H.~L.} \emph{et~al.}
\newblock \bibinfo{title}{qubit-adapt-vqe: An adaptive algorithm for
  constructing hardware-efficient ans{\"a}tze on a quantum processor}.
\newblock \emph{\bibinfo{journal}{PRX Quantum}} \textbf{\bibinfo{volume}{2}},
  \bibinfo{pages}{020310} (\bibinfo{year}{2021}).

\bibitem{mcclean_barren_2018}
\bibinfo{author}{McClean, J.~R.}, \bibinfo{author}{Boixo, S.},
  \bibinfo{author}{Smelyanskiy, V.~N.}, \bibinfo{author}{Babbush, R.} \&
  \bibinfo{author}{Neven, H.}
\newblock \bibinfo{title}{Barren plateaus in quantum neural network training
  landscapes}.
\newblock \emph{\bibinfo{journal}{Nat. Commun.}}
  \textbf{\bibinfo{volume}{9}}, \bibinfo{pages}{1--6} (\bibinfo{year}{2018}).

\bibitem{cerezo_cost-function-dependent_2020}
\bibinfo{author}{Cerezo, M.}, \bibinfo{author}{Sone, A.},
  \bibinfo{author}{Volkoff, T.}, \bibinfo{author}{Cincio, L.} \&
  \bibinfo{author}{Coles, P.~J.}
\newblock \bibinfo{title}{Cost function dependent barren plateaus in shallow
  parametrized quantum circuits}.
\newblock \emph{\bibinfo{journal}{Nat. Commun.}}
  \textbf{\bibinfo{volume}{12}}, \bibinfo{pages}{1--12} (\bibinfo{year}{2021}).

\bibitem{cruise2020practical}
\bibinfo{author}{Cruise, J.~R.}, \bibinfo{author}{Gillespie, N.~I.} \&
  \bibinfo{author}{Reid, B.}
\newblock \bibinfo{title}{Practical quantum computing: The value of local
  computation}.
\newblock \emph{\bibinfo{journal}{arXiv preprint arXiv:2009.08513}}
  (\bibinfo{year}{2020}).

\bibitem{zhang2020collective}
\bibinfo{author}{Zhang, D.-B.} \& \bibinfo{author}{Yin, T.}
\newblock \bibinfo{title}{Collective optimization for variational quantum
  eigensolvers}.
\newblock \emph{\bibinfo{journal}{Phys. Rev. A}}
  \textbf{\bibinfo{volume}{101}}, \bibinfo{pages}{032311}
  (\bibinfo{year}{2020}).

\bibitem{ibmq-devices}
\bibinfo{note}{\emph{ibmq\_manhattan} (v1.5.1), \emph{ibmq\_toronto} (v1.1.4),
  \emph{ibmq\_paris} (v1.6.5), \emph{ibmq\_santiago} (v1.3.0),
  \emph{ibmq\_athens} (v1.3.3), \emph{ibmq\_valencia} (v1.4.3). IBM Quantum
  team. Retrieved from \url{https://quantum-computing.ibm.com} (2020).}

\bibitem{peruzzo_variational_2014}
\bibinfo{author}{Peruzzo, A.} \emph{et~al.}
\newblock \bibinfo{title}{A variational eigenvalue solver on a photonic quantum
  processor}.
\newblock \emph{\bibinfo{journal}{Nat. Commun.}}
  \textbf{\bibinfo{volume}{5}}, \bibinfo{pages}{1--7} (\bibinfo{year}{2014}).

\bibitem{frazier_tutorial_2018}
\bibinfo{author}{Frazier, P.~I.}
\newblock \bibinfo{title}{A tutorial on bayesian optimization}.
\newblock \emph{\bibinfo{journal}{arXiv preprint arXiv:1807.02811}}
  (\bibinfo{year}{2018}).

\bibitem{wigley2016fast}
\bibinfo{author}{Wigley, P.~B.} \emph{et~al.}
\newblock \bibinfo{title}{Fast machine-learning online optimization of
  ultra-cold-atom experiments}.
\newblock \emph{\bibinfo{journal}{Sci. Rep.}}
  \textbf{\bibinfo{volume}{6}}, \bibinfo{pages}{1--6} (\bibinfo{year}{2016}).

\bibitem{nakamura2019non}
\bibinfo{author}{Nakamura, I.}, \bibinfo{author}{Kanemura, A.},
  \bibinfo{author}{Nakaso, T.}, \bibinfo{author}{Yamamoto, R.} \&
  \bibinfo{author}{Fukuhara, T.}
\newblock \bibinfo{title}{Non-standard trajectories found by machine learning
  for evaporative cooling of 87 rb atoms}.
\newblock \emph{\bibinfo{journal}{Opt. Express}}
  \textbf{\bibinfo{volume}{27}}, \bibinfo{pages}{20435--20443}
  (\bibinfo{year}{2019}).

\bibitem{mukherjee_bayesian_2020}
\bibinfo{author}{Mukherjee, R.}, \bibinfo{author}{Xie, H.} \&
  \bibinfo{author}{Mintert, F.}
\newblock \bibinfo{title}{Bayesian optimal control of
  greenberger-horne-zeilinger states in rydberg lattices}.
\newblock \emph{\bibinfo{journal}{Phys. Rev. Lett.}}
  \textbf{\bibinfo{volume}{125}}, \bibinfo{pages}{203603}
  (\bibinfo{year}{2020}).

\bibitem{sauvage_optimal_2020}
\bibinfo{author}{Sauvage, F.} \& \bibinfo{author}{Mintert, F.}
\newblock \bibinfo{title}{Optimal quantum control with poor statistics}.
\newblock \emph{\bibinfo{journal}{PRX Quantum}} \textbf{\bibinfo{volume}{1}},
  \bibinfo{pages}{020322} (\bibinfo{year}{2020}).

\bibitem{otterbach2017unsupervised}
\bibinfo{author}{Otterbach, J.} \emph{et~al.}
\newblock \bibinfo{title}{Unsupervised machine learning on a hybrid quantum
  computer}.
\newblock \emph{\bibinfo{journal}{arXiv preprint arXiv:1712.05771}}
  (\bibinfo{year}{2017}).

\bibitem{zhu2019training}
\bibinfo{author}{Zhu, D.} \emph{et~al.}
\newblock \bibinfo{title}{Training of quantum circuits on a hybrid quantum
  computer}.
\newblock \emph{\bibinfo{journal}{Sci. Adv.}}
  \textbf{\bibinfo{volume}{5}}, \bibinfo{pages}{eaaw9918}
  (\bibinfo{year}{2019}).

\bibitem{sung2020using}
\bibinfo{author}{Sung, K.~J.} \emph{et~al.}
\newblock \bibinfo{title}{Using models to improve optimizers for variational
  quantum algorithms}.
\newblock \emph{\bibinfo{journal}{Quantum Sci. Technol.}}
  \textbf{\bibinfo{volume}{5}}, \bibinfo{pages}{044008} (\bibinfo{year}{2020}).

\bibitem{ovchinnikov2003antiferromagnetic}
\bibinfo{author}{Ovchinnikov, A.}, \bibinfo{author}{Dmitriev, D.},
  \bibinfo{author}{Krivnov, V.~Y.} \& \bibinfo{author}{Cheranovskii, V.}
\newblock \bibinfo{title}{Antiferromagnetic ising chain in a mixed transverse
  and longitudinal magnetic field}.
\newblock \emph{\bibinfo{journal}{Phys. Rev. B}}
  \textbf{\bibinfo{volume}{68}}, \bibinfo{pages}{214406}
  (\bibinfo{year}{2003}).

\bibitem{gray_quimb_2018}
\bibinfo{author}{Gray, J.}
\newblock \bibinfo{title}{quimb: A python package for quantum information and
  many-body calculations}.
\newblock \emph{\bibinfo{journal}{J. Open Source Softw.}}
  \textbf{\bibinfo{volume}{3}}, \bibinfo{pages}{819} (\bibinfo{year}{2018}).

\bibitem{ibmq-qiskit}
\bibinfo{author}{Abraham, H.} \emph{et~al.}
\newblock \bibinfo{title}{Qiskit: An open-source framework for quantum
  computing} (\bibinfo{year}{2019}).
\newblock \urlprefix\url{https://doi.org/10.5281/zenodo.2562110}.

\bibitem{Bravyi2017Jan}
\bibinfo{author}{Bravyi, S.}, \bibinfo{author}{Gambetta, J.~M.},
  \bibinfo{author}{Mezzacapo, A.} \& \bibinfo{author}{Temme, K.}
\newblock \bibinfo{title}{Tapering off qubits to simulate fermionic
  hamiltonians}.
\newblock \emph{\bibinfo{journal}{arXiv preprint arXiv:1701.08213}}
  (\bibinfo{year}{2017}).

\bibitem{openfermion}
\bibinfo{author}{McClean, J.~R.} \emph{et~al.}
\newblock \bibinfo{title}{Openfermion: the electronic structure package for
  quantum computers}.
\newblock \emph{\bibinfo{journal}{Quantum Sci. Technol.}}
  \textbf{\bibinfo{volume}{5}}, \bibinfo{pages}{034014} (\bibinfo{year}{2020}).

\bibitem{bridgeman_tensor-networks_2017}
\bibinfo{author}{Bridgeman, J.~C.} \& \bibinfo{author}{Chubb, C.~T.}
\newblock \bibinfo{title}{Hand-waving and interpretive dance: an introductory
  course on tensor networks}.
\newblock \emph{\bibinfo{journal}{J. Phys. A}} \textbf{\bibinfo{volume}{50}}, \bibinfo{pages}{223001}
  (\bibinfo{year}{2017}).

\bibitem{tensorflow_2015}
\bibinfo{author}{Abadi, M.} \emph{et~al.}
\newblock \bibinfo{title}{{TensorFlow}: Large-scale machine learning on
  heterogeneous systems} (\bibinfo{year}{2015}).
\newblock \urlprefix\url{https://www.tensorflow.org/}.
\newblock \bibinfo{note}{Software available from tensorflow.org}.

\bibitem{gpyopt2016}
\bibinfo{author}{Authors}.
\newblock \bibinfo{title}{Gpyopt: A bayesian optimization framework in python}.
\newblock \bibinfo{howpublished}{\url{http://github.com/SheffieldML/GPyOpt}}
  (\bibinfo{year}{2016}).

\bibitem{snoek2012practical}
\bibinfo{author}{Snoek, J.}, \bibinfo{author}{Larochelle, H.} \&
  \bibinfo{author}{Adams, R.~P.}
\newblock \bibinfo{title}{Practical bayesian optimization of machine learning
  algorithms}.
\newblock \emph{\bibinfo{journal}{Adv. Neural Inf. Process Syst.}} \textbf{\bibinfo{volume}{25}}, \bibinfo{pages}{2951--2959}
  (\bibinfo{year}{2012}).

\bibitem{shahriari2015taking}
\bibinfo{author}{Shahriari, B.}, \bibinfo{author}{Swersky, K.},
  \bibinfo{author}{Wang, Z.}, \bibinfo{author}{Adams, R.~P.} \&
  \bibinfo{author}{De~Freitas, N.}
\newblock \bibinfo{title}{Taking the human out of the loop: A review of
  bayesian optimization}.
\newblock \emph{\bibinfo{journal}{Proc. IEEE}}
  \textbf{\bibinfo{volume}{104}}, \bibinfo{pages}{148--175}
  (\bibinfo{year}{2015}).

\bibitem{bartlett_alternative_1989}
\bibinfo{author}{Bartlett, R.~J.}, \bibinfo{author}{Kucharski, S.~A.} \&
  \bibinfo{author}{Noga, J.}
\newblock \bibinfo{title}{Alternative coupled-cluster ans{\"a}tze ii. the
  unitary coupled-cluster method}.
\newblock \emph{\bibinfo{journal}{Chem. Phys. Lett.}}
  \textbf{\bibinfo{volume}{155}}, \bibinfo{pages}{133--140}
  (\bibinfo{year}{1989}).

\bibitem{yeter2021benchmarking}
\bibinfo{author}{Yeter-Aydeniz, K.} \emph{et~al.}
\newblock \bibinfo{title}{Benchmarking quantum chemistry computations with
  variational, imaginary time evolution, and krylov space solver algorithms}.
\newblock \emph{\bibinfo{journal}{arXiv preprint arXiv:2102.05511}}
  (\bibinfo{year}{2021}).

\bibitem{imperial-rcs-acknowledgement}
\bibinfo{title}{Imperial college research computing service}.
\newblock \urlprefix\url{https://doi.org/10.14469/hpc/2232}.

\end{thebibliography}

\end{document}